\documentclass{aa}
\usepackage{graphicx}
\usepackage{txfonts}
\begin{document}
   \title{A model for selective line-driven acceleration of \\ ions in the solar 
winds of OB-type stars \\ occurring via the nonlinear process \\ of stimulated 
Rayleigh scattering}

   \author{P. P. Sorokin\inst{1}
          \and J. H. Glownia\inst{2}}

   \institute{IBM Research Division, P. O. Box 218, Yorktown Heights, NY  10598-0218, USA\\
              \email{sorokin@us.ibm.com}
          \and MS J585, Los Alamos National Laboratory, P. O. Box 1663,
           Los Alamos, NM  87545-1663, USA\\
           \email{jglownia@lanl.gov}}

   \date{Received 2004; accepted 2004}

   \abstract{A conceptually novel model - one in which stimulated 
($i.e.$ induced) resonance Rayleigh scattering causes ions of select species to 
become accelerated to very high terminal velocities in the solar winds of 
OB-type stars - is proposed to explain the general appearance of so-called P 
Cygni profiles that often dominate the vacuum ultraviolet (VUV) spectra of 
such stars. In the unit step of the proposed scattering process, a radially 
outgoing photon (frequency $\nu _1 )$ that is arbitrarily blueshifted with 
respect to the rest frame frequency $\nu _o $ of the resonance line 
associated with a P Cygni profile is absorbed from the illuminating star's 
continuum, a new photon (frequency $\nu _2 )$ that is very close to $\nu _o 
$ and that propagates in the \textit{backwards }direction ($i. e.$ towards the star) is created, and 
the outwardly directed velocity $v$ of the ion being accelerated is increased 
by an amount $\Delta v\approx 2h\nu _o /cm_I ,$ with all three events 
occurring simultaneously. A monochromatic wave at $\nu _2 $ initially forms 
at some distance from the star, and becomes enormously amplified via the 
stimulated scattering process as it propagates radially inwards towards the 
star, all the while retaining a relatively high degree of monochromaticity. 
The high rate of stimulated scattering enables ions of the resonant species 
in the solar wind to become accelerated to terminal velocities as high as 
$\sim $1000 km/sec. Since stimulated scattering processes are characterized 
by pump power thresholds, the model readily explains why only select species 
are accelerated to very high velocities in a given star's solar wind, and 
also why a dramatic P Cygni profile for a given ion species can often 
discontinuously be present or absent when spectra of stars varying only 
slightly in spectral type are compared. 
   
   }

\titlerunning{A model for selective line-driven acceleration of ions}
\authorrunning{P. P. Sorokin \& J. H. Glownia}
   \maketitle
%

\section{Introduction}

So-called P Cygni profiles of select ion resonances ($e.g.$ C IV $\lambda 
\lambda $1548, 1551; Si IV $\lambda \lambda$1394, 1403; NV $\lambda 
\lambda $1239, 1243 -- see Fig. 1) are often the dominant features in the 
vacuum ultraviolet (VUV) spectra of O-type (Walborn $et$ $al.$ \cite{walborn85}) and B-type 
(Walborn $et$ $al.$ \cite{walborn95}) stars, especially giants and supergiants. These 
features were originally discovered in the late nineteen-sixties by Morton 
(Morton \cite{morton67}) and his colleagues (Morton $et$ $al.$ \cite{morton69}) with the use of 
rocket-based instrumentation, and have since been studied extensively by 
many astronomers utilizing several orbiting satellites that were 
successfully launched in the years that followed. Especially productive in 
obtaining P Cygni spectral data were the fully dedicated \textit{Copernicus} and \textit{International Ultraviolet Explorer (IUE)} satellites, 
and the \textit{Space Telescope Imaging Spectrometer (STIS)} aboard the \textit{Hubble Space Telescope.} 

\begin{figure}
\resizebox{\hsize}{!}{\includegraphics{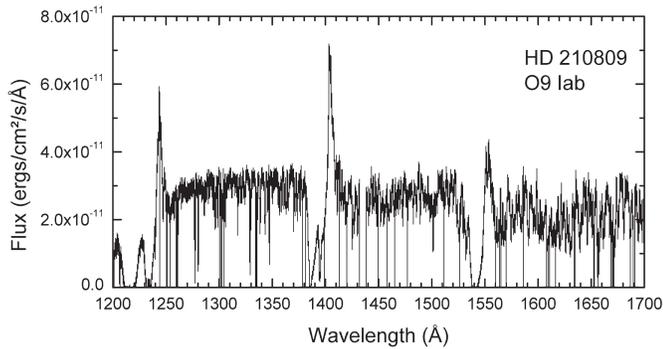}}
\caption{VUV spectrum of the O9 supergiant HD 210809 recorded with 
the \textit{Space Telescope Imaging Spectrometer (STIS) }aboard the \textit{Hubble Space Telescope}. (Spectrum downloaded from the \textit{MAST Scrapbook }interactive web site.) The 
three P Cygni profiles dominating the spectrum are (from left to right) due 
to N V, Si IV, and C IV ions, respectively. The absorption at 1216 {\AA} is 
that of H-atom Lyman alpha. A multitude of narrow interstellar absorption 
lines are present. Wider stellar photospheric absorptions can also be seen. }
\label{fig1}
\end{figure}

In a typical P Cygni profile, the star's continuum level in a relatively 
broad spectral region that is \textit{blueshifted} with respect to a given ion resonance line 
$\nu _o $ is heavily absorbed. (Frequently the ion resonances are doublets, 
as is the case with the three prominent P Cygni profiles shown in Fig.~\ref{fig1}.) 
The measured frequency displacement from $\nu _o $ of the short wavelength 
limit $\nu _B $ of this absorbed region is normally used by astronomers to 
determine the maximum velocity $v_{\max } $ of the corresponding ions in the 
solar wind via the standard relationship $\nu _B -\nu _o =\nu _o v_{\max } 
/c.$ From such Doppler-shift-based analyses of P Cygni profiles, astronomers 
have inferred that the corresponding ions are strongly accelerated radially 
away from the stars, reaching in some cases maximum velocities as high as 
two or three thousand km/sec. For a hot star, the maximum velocity deduced 
in this manner is frequently greater than the calculated escape velocity 
$V_e =\left( {2GM/R} \right)^{1/2}$ at the radius $R$ of the photosphere, 
implying an actual ejection of matter from the star into interstellar space.

Included in Morton (\cite{morton76}) are special \textit{Copernicus }VUV spectral scans of the O4 If 
supergiant $\zeta $ Pup, recorded at very high spectral resolution (0.051 
{\AA} FWHM), and with an exceptionally high signal-to-noise ratio. For this 
very hot star, which has $R/R_\odot \approx $ 20.3 and $M/M_\odot \approx 
$ 100, it was calculated in Morton (1976) that $V_e =$1370 $\pm $ 160 km/sec. 
Between 920 {\AA} and 1750 {\AA}, 13 different ion species were found to 
display P Cygni profiles in $\zeta $ Pup, many of them appearing very 
strong. For example, the ion C III was observed to have a strong P Cygni 
profile associated with its resonance line at $\approx $ 977 {\AA}. The blue 
edge of the heavily absorbed region appeared shifted from 977 {\AA} by 
$\approx $ 2710 km/sec, prompting the conclusion that was made in Morton 
(1976) that C III ions in the solar wind of $\zeta $ Pup are continually 
being ejected into space. However, as will be shortly pointed out, according 
to the nonlinear P Cygni model here being proposed, use of the relationship 
$\nu _B -\nu _o =\nu _o v_{\max } /c$ to determine the maximum ion velocity 
$v_{\max } $ attained in a star's solar wind results in an overestimation of 
this quantity by almost exactly a factor 2. 

As noted in Lamers {\&} Cassinelli (\cite{lamers}), current stellar wind theories 
fall into three broad classes: radiative ($i.e.$ line-driven) models, coronal 
models, and hybrid models. In radiative models, transfer of photon momentum 
to the gas is assumed to occur through the opacity of the many strong VUV 
resonance lines that are present. Increased acceleration is believed to 
result from the progressive Doppler shifting of the line opacity into the 
unattenuated photospheric radiation field. The stellar wind in existing 
radiative models is thus assumed to be driven by a purely \textit{linear} effect, commonly 
known as \textit{radiation pressure}. 

In the model here being proposed, a \textit{nonlinear} photomechanism drives the acceleration 
of the fast moving ions in the solar wind, simultaneously producing the P 
Cygni spectral profiles one sees in the line-of-sight to the star. Perhaps 
the most attractive feature of such a nonlinear mechanism would be that it 
would possess a definite pump power threshold. Only those ion species for 
which this threshold is reached would be accelerated to very high velocities 
in a given star's solar wind. Such selectivity is very hard to explain with 
linear radiative models for solar winds. 

In the present paper the focus is placed upon the basic physics of the 
proposed nonlinear model. No attempt is made to account in detail for 
fine-grained features appearing in P Cygni profiles of specific stars. The 
paper is organized as follows. In Sect. 2, C II $\lambda \lambda $1334, 
1336 ion resonance spectra recorded in four different stars of roughly the 
same spectral type are compared to demonstrate the abrupt manner in which P 
Cygni profiles often appear in star spectra, thus strongly suggesting that a 
threshold effect of some kind must be involved. In Sect. 3, the proposed 
nonlinear scattering mechanism is outlined. Undoubtedly of some significance 
is the fact that this mechanism is easily shown to be consistent with the 
conservation of both energy and momentum. Simple consideration of the manner 
in which spontaneous resonance Rayleigh scattering ($i.e.$ elastic scattering) 
would occur in the rest frame of an ion being accelerated in a star's solar 
wind directly leads to identification of the proposed nonlinear scattering 
mechanism with stimulated ($i.e.$ induced) resonance Rayleigh scattering. In the 
unit step of this stimulated process, three distinct events \textit{simultaneously} occur. (\ref{eq1}) An 
outwardly propagating photon from the star's continuum that is blueshifted 
with respect to the rest frame frequency $\nu _o $ of a strong ion resonance 
line becomes absorbed by the ion. (\ref{eq2}) A photon at exactly $\nu _o $ that 
propagates radially backwards ($i. e. $ towards the star) is emitted by the ion, 
contributing to the intensity of a backward propagating light wave at $\nu 
_o $ that is already present at the position of the ion. (For analytical 
purposes, this wave is termed the $\nu _2 $ wave, but its frequency is 
$\approx \nu _o .)$ (\ref{eq3}) The outwardly directed radial velocity $v$ of the 
ion is increased by $\Delta v\approx 2h\nu _o /cm_I $, \textit{irrespective} of the value of $v$. 

Since, at any point in the solar wind, the transition probability for the 
unit step to occur is proportional to the product of the outwardly 
propagating continuum flux and the inwardly propagating light beam intensity 
$I_2 $, the likely occurrence of a stimulated scattering regime is therefore 
here expected, with the $\nu _2 $ wave continuously gaining in intensity as 
it propagates inwardly towards the star. During the entire time it undergoes 
amplification, the $\nu _2 $ wave remains relatively monochromatic, a result 
of the unit step in the scattering process becoming stimulated by the same 
wave. However, as the $\nu _2 $ wave propagates inwardly towards the star, 
the frequency of the pump light that effectively drives the stimulated 
scattering process continually changes. In this way, continuum light from 
the star spanning a broad spectral range can contribute to the intensity of 
the monochromatic $\nu _2 $ wave as the latter impinges upon the photosphere 
of the star. 

In Sect. 4, equations that should describe the stimulated scattering process 
are briefly discussed. However, efforts to provide numerical solutions for 
these equations have not yet been attempted. 

In the next section of the present paper (Sect. 5), two types of 
spectroscopic evidence that could help to substantiate the nonlinear model 
are discussed -- fluorescence spectra and two-photon absorption bands. Most 
P Cygni profiles are observed to possess a prominent fluorescence component. 
Via simple extension of the stimulated Rayleigh scattering model used to 
explain P Cygni absorption, one can also comprehend how P Cygni fluorescence 
is excited. By considering the effect of having the absorption and the 
fluorescence excitation mechanisms operate in tandem, one can then easily 
account for the observed fact that the strongest apparent absorption in a P 
Cygni profile occurs at frequencies considerably blueshifted from $\nu _o .$ 

In Sect. 5 it is also suggested that one might seek to identify Doppler 
broadened absorption bands representing resonantly enhanced two-photon 
absorption of outwardly propagating continuum light from a star by various 
ion species present in its photosphere, or just outside. Such two-photon 
absorption could in principle be induced by the powerful, inwardly 
propagating, monochromatic beams at the P Cygni ion rest frame frequencies 
$\nu _o $, provided that near resonances exist between some of the 
frequencies $\nu _o $ and some transitions of the two-photon-absorbing ions. 
Details of a search for such near resonances in the case of the P Cygni star 
$\zeta $ Pup are described. However, with one possible intriguing exception, 
this search did not result in the clear identification of any two-photon 
absorption bands in this star. 

In Sect. 6 nonlinear and linear P Cygni mechanisms are briefly compared. 

\section{P Cygni profiles in stars of almost similar spectral types}

\begin{figure}
\resizebox{\hsize}{!}{\includegraphics{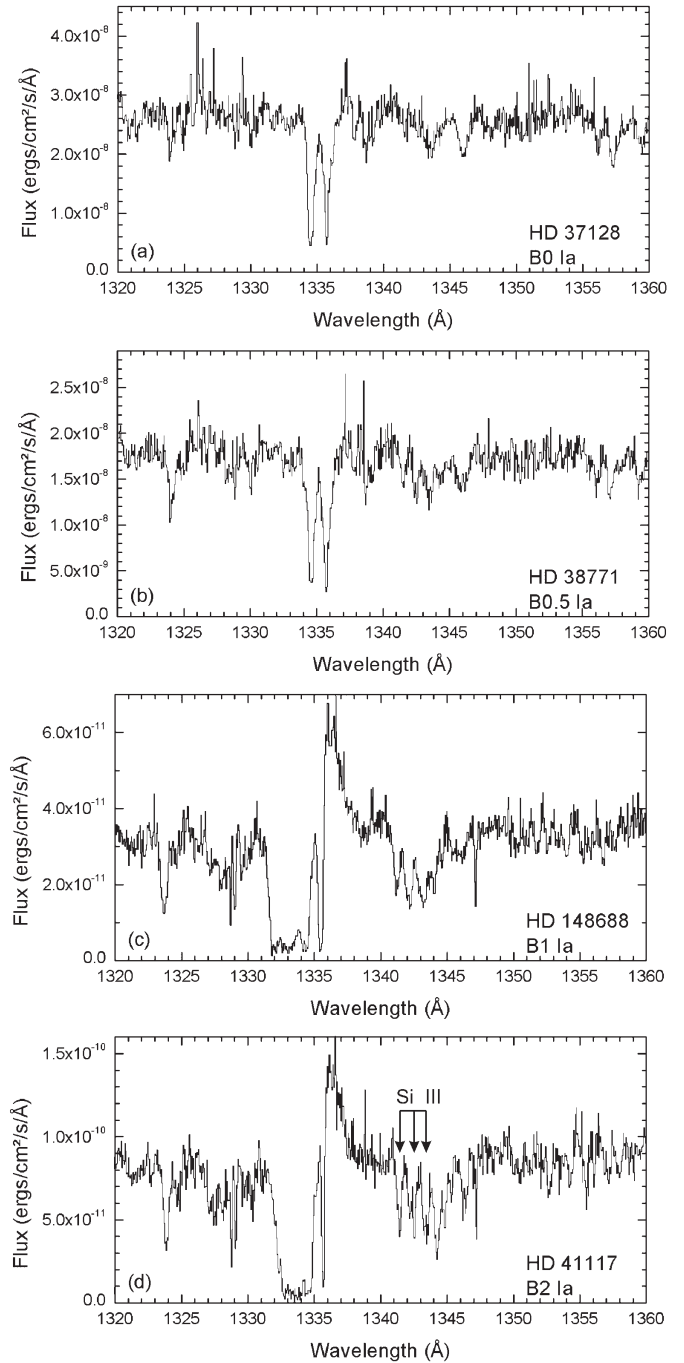}}
\caption{Montage of VUV spectra of B-type supergiants recorded with 
the \textit{International Ultraviolet Explorer (IUE).}(Spectra downloaded from the \textit{MAST Scrapbook}
interactive web site.) The spectral region 
shown here includes the C II $\lambda \lambda $1335, 1336 resonance 
doublet. Stars in adjacent spectra are relatively close in temperature.}
\label{fig2}
\end{figure}

In Fig.~\ref{fig2}, VUV spectra recorded with the \textit{IUE }satellite of four B-type 
supergiants are displayed over a limited wavelength range that includes the 
strongly allowed C II ion resonance lines at 1334.5 {\AA} and 1335.7 {\AA}. 
Each star shown is of a different spectral type, but the difference in 
temperatures between stars in adjacent spectra is relatively small. In the 
figure, star temperatures decrease in going from top to bottom. 

It is seen that in (a) and (b) the C II doublet appears normally absorbing. 
Whether the absorbing ions in these two spectra are located in the stars' 
photospheres, or just outside, is somewhat difficult to judge. It may also 
be that the lines, at least in part, represent interstellar absorptions. 
Whatever the case, a significant spectral change is seen to occur in (c), 
namely, a P Cygni profile, with its characteristic blueshifted region of 
strong absorption, dramatically appears. From all four spectra, it is clear 
that this intense new absorption is not simply a broad feature of the 
stellar photosphere. To explain its sudden appearance via theories based 
upon the effects of linear radiation pressure also seems excessively 
difficult. Linear photoexcitation is probably entirely responsible for the 
comparatively narrow C II absorption lines seen in (a) and (b). These lines 
are still seen to be present in (c) and (d), and linear absorption may still 
be the agent that produces them. However, it is quite evident from the 
spectral sequence shown in Fig.~\ref{fig2} that a dramatic new mechanism 
significantly increasing the degree to which continuum light emitted by a 
star is attenuated can sometimes abruptly become activated. In the present 
paper it is proposed that this mechanism is stimulated Rayleigh scattering. 
The rationale for making such a hypothesis is presented throughout the remainder of the 
paper. 

\section{Identifying the proposed P Cygni mechanism with stimulated resonance 
Rayleigh scattering}

In any type of stimulated scattering mechanism, one would expect both 
momentum and energy to be conserved in the unit step of the process. As 
already outlined in the Introduction, in the unit step of the process here 
being invoked to explain P Cygni profiles, three events simultaneously occur 
at the position of an ion being accelerated in the solar wind. (\ref{eq1}) A 
continuum photon propagating radially outwards from the star, and having an 
arbitrary frequency $\nu _1 $ that is blueshifted with respect to the rest 
frame frequency $\nu _o $ of a resonance transition of the ion, is absorbed 
by the latter. (\ref{eq2}) A photon of frequency $\nu _2 $ is emitted by the ion in 
the direction pointing radially inwards towards the star. (\ref{eq3}) The radial 
velocity of the ion is increased from $v$ to $v+\Delta v.$ 

Let relativistic corrections initially be neglected. Requiring that energy 
be conserved in the unit scattering step implies that
\begin{equation}
\label{eq1}
h\nu _1 \cong h\nu _2 +m_I v(\Delta v),
\end{equation}
while requiring that momentum be conserved dictates that
\begin{equation}
\label{eq2}
\frac{h\nu _1 }{c}+\frac{h\nu _2 }{c}=m_I (\Delta v).
\end{equation}
From these equations, it then follows that
\begin{equation}
\label{eq3}
\nu _1 =\nu _2 \frac{\left( {1+\frac{v}{c}} \right)}{\left( {1-\frac{v}{c}} 
\right)}.
\end{equation}
Equation (\ref{eq3}) implies that energy and momentum could in principle both be 
conserved, even if the value of $\nu _2 $ were to differ in each scattering 
event. However, if the photon emitted at $\nu _2 $ were always at the same 
fixed frequency, this would allow the $\nu _2 $ light wave to become 
amplified much more as it propagates radially inwards towards the star, 
since the rate of stimulated scattering at any point in the solar wind is 
proportional to the $\nu _2 $ light wave intensity at that same point. The 
latter would obviously be much greater if the backwards-emitted photon in 
each unit step were always at the same fixed frequency. In the proposed 
model, not only is this assumed to be the case, but it is also postulated 
that $\nu _2 \equiv \nu _o .$

One can also attempt to view the unit step in the proposed scattering 
process simply as spontaneous resonance Rayleigh scattering ($i.e. $ elastic 
scattering) occurring in the rest frame of the ion being accelerated. Viewed 
in this rest frame, the pumping frequency ($i. e. $ the incident continuum photon 
frequency) is $\nu _1 -\nu _1 (v/c),$ since the ion is receding from the 
star at velocity $v.$ (Here again, relativistic corrections are being 
ignored.) Thus, in the ion rest frame, an oscillating dipole moment would be 
induced at the same frequency, $\nu _1 -\nu _1 (v/c).$ This oscillating 
dipole moment would radiate in all directions. Photons that are emitted in 
the backwards direction ($i. e. $ towards the star) would therefore be at $\nu _2 
=\left[ {\nu _1 -\nu _1 (v/c)} \right]-\left[ {\nu _1 -\nu _1 (v/c)} 
\right](v/c),$ implying that 
\begin{equation}
\label{eq4}
\nu _1 =\frac{\nu _2 }{\left[ {1-(v/c)} \right]^2}.
\end{equation}
The values of $\nu _1 $ given by Eqs. (\ref{eq3}) and (\ref{eq4}) are the same to within a 
factor $\left[ {1-(v^2/c^2)} \right],$ which would result in a difference 
between these quantities of only about one wavenumber at typical resonance 
frequencies ($\sim $ 100,000 cm$^{-1})$ of solar wind ions moving at $\sim 
$ 1000 km/sec. Although the effects of this difference on the proposed P 
Cygni scattering model would be very small, it is nonetheless of some 
interest to compare the two scenarios when relativistic effects are taken 
into account. We here consider first the changes that would occur in the 
spontaneous Rayleigh scattering picture. Use of the known expression for the 
relativistic Doppler shift would imply that, in the rest frame of an ion in 
the solar wind moving away from the star with velocity $v,$ the frequency of 
an incident continuum photon at $\nu _1 $ would be 
\begin{equation}
\label{eq5}
{\nu }'=\nu _1 \frac{\sqrt {1-\frac{v^2}{c^2}} }{\left( {1+\frac{v}{c}} 
\right)}=\nu _1 \sqrt {\frac{1-\frac{v}{c}}{1+\frac{v}{c}}} .
\end{equation}
Therefore, the frequency of photons emitted in the direction of the star by 
a dipole moment oscillating in the ion rest frame at the frequency ${\nu }'$ 
would be 
\[
\nu _2 ={\nu }'\sqrt {\frac{1-\frac{v}{c}}{1+\frac{v}{c}}} =\nu _1 
\frac{\left( {1-\frac{v}{c}} \right)}{\left( {1+\frac{v}{c}} \right)}.
\]
Interestingly enough, the above equation expresses exactly the same 
relationship between $\nu _1 $ and $\nu _2 $ that is represented in Eq. (\ref{eq3}). 
To complete the comparison of scenarios, one should finally consider the 
equations for conservation of energy and momentum that would hold in the 
unit step of the proposed scattering process when relativistic effects are 
included. The equation equivalent to Eq. (\ref{eq1}) is:
\begin{equation}
\label{eq6}
h\nu _1 \cong h\nu _2 +m_I v\left( {1-\frac{v^2}{c^2}} \right)^{-3/2}(\Delta 
v),
\end{equation}
while that equivalent to Eq. (\ref{eq2}) is:
\begin{equation}
\label{eq7}
\frac{h\nu _1 }{c}+\frac{h\nu _2 }{c}=m_I \left( {1-\frac{v^2}{c^2}} 
\right)^{-3/2}(\Delta v).
\end{equation}
From Eqs. (\ref{eq6}) and (\ref{eq7}), an equation identical to Eq. (\ref{eq3}) follows. Thus, when 
relativistic effects are included, the equations relating $\nu _1 $ and $\nu 
_2 $ in the unit step of the proposed scattering process and in spontaneous 
Rayleigh scattering are seen to be identical, strongly implying that the 
former should be identified with the latter. As has already been emphasized, 
in the proposed model it is assumed that $\nu _2 \equiv \nu _o ,$ and that 
the presence of an intense light wave at this same frequency at the position 
of every ion in the solar wind \textit{preferentially stimulates} ($i. e. $ induces) the photons produced by the 
Rayleigh scattering process to be emitted in the backwards direction, while 
simultaneously causing the ion to be accelerated radially outwards. \textit{While it is hard to provide the conditions necessary to unambiguously demonstrate stimulated Rayleigh scattering in the laboratory, it is nonetheless quite apparent that the astrophysical environments of hot stars which display P Cygni profiles would offer optimum conditions for this particular type of stimulated scattering to occur. }

In the model, all photons produced by the stimulated Rayleigh scattering 
process throughout the entire volume occupied by the solar wind occur at the 
same monochromatic frequency $\nu _o $ and propagate radially inwards 
towards the star. All photons nonlinearly absorbed by the stimulated 
scattering process are continuum photons that are emitted from the 
photosphere of the illuminating star and propagate radially away from it. 
The frequencies of these absorbed photons, which at all points in the solar 
wind supply the pumping energy needed to drive the stimulated Rayleigh 
scattering process, span a wide spectral range that originates at $\nu _o 
$ and extends to higher energies. From Eq. (\ref{eq3}), an ion moving with radial 
velocity $v$ will nonlinearly absorb continuum light in a narrow frequency 
band centered at $\nu _1 =\nu _o \left( {1+\frac{v}{c}} \right)\left( 
{1-\frac{v}{c}} \right)^{-1}\simeq \nu _o +2\nu _o (v/c).$ As an ion becomes 
accelerated more and more, the continuum light it effectively absorbs occurs 
at higher and higher frequencies. One thus can see how the stimulated 
Rayleigh scattering process is able to efficiently convert most of the 
incoherent light continuum photons emitted from a star over a wide, 
blueshifted spectral range into the same number of essentially monochromatic 
coherent light photons in the form of a spherical light wave at $\nu _o $ 
that radially converges upon the star's photosphere. 

From the above paragraph, one can now comprehend the basis for the statement 
made in the Introduction that use of the standard relationship $\nu _B -\nu 
_o =\nu _o v_{\max } /c$ to determine the maximum ion velocity $v_{\max } $ 
attained in a star's solar wind results in an overestimation of this 
quantity by almost exactly a factor 2. However, this statement does assume 
that the most blueshifted absorption occurring in a P Cygni band almost 
entirely represents \textit{nonlinear }absorption.   

From Eq. (\ref{eq7}) one has that 
\begin{equation}
\label{eq8}
\Delta v\simeq \frac{2h\nu _o }{cm_I }.
\end{equation}
This equation states that the velocity increase $\Delta v$ occurring in each 
unit scattering event is always roughly the same - that is, to a first 
approximation, $\Delta v$ is not a function of the velocity $v$ of the ion 
involved in the scattering event. It depends only on inherent properties of 
the ion being accelerated in the solar wind. In the case of the C II ion P 
Cygni profiles shown in the two lower spectra of Fig.~\ref{fig2}, one has that 
$\Delta v\simeq $ 50 cm/sec. Although each nonlinear scattering event results 
in only a modest velocity increase for the ion being accelerated, the rate 
of occurrence of such events will be very large in the region of the solar 
wind where the particles are being significantly accelerated, due to the 
Rayleigh scattering process becoming stimulated. In the case of Fig.~\ref{fig2}c, the 
short wavelength limit of the P Cygni absorption region appears offset from 
the shortest-wavelength C II doublet component by about 200 cm$^{-1}$. Thus, 
according to the proposed nonlinear model, $v_{\max } $ would here be about 
395 km/sec. An ion accelerated to the maximum velocity $v_{\max } $ in the 
solar wind would therefore have had to participate in at least $\sim 
$ 790,000 unit step scatterings. 

\section{Outline of a model for stimulated Rayleigh scattering occurring 
in P Cygni stars} 

In principle, one should be able to model the proposed stimulated scattering 
scenario with a set of differential equations involving a number of 
dependent and independent variables. Equations for a model possessing 
minimum mathematical complexity are outlined below, and some discussion is 
given of the parameters involved. However, at this stage only qualitative 
statements regarding this model can be made, as attempts have not yet been 
made to provide computer-based numerical solutions for even this rudimentary 
system of equations. 

In the conceptually simplest type of model, there would be only one 
independent variable. This would be $r,$ the radial distance from the center 
of the illuminating star. One primary dependent variable would be $v(r),$ 
the velocity of an ion in the solar wind at radial position $r.$ Assuming 
that there is only a single ion velocity associated with each value of 
$r$ effectively presupposes that the nonlinear ion acceleration process 
($i. e. $ stimulated Rayleigh scattering) commences at some given radius $R_o >R$ 
($R$ being the star's photospheric radius), and that at this radius a 
spherically uniform, radially expanding flow of ions occurs, with each ion 
in the flow moving at the same radial velocity $v(R_o ).$ Let $K_2 $ 
represent the total rate at which ions of a given P Cygni species 
continually cross the surface of the sphere at $R_o ,$ $ i.e.$ $K_2 =4\pi R_o^2 n_I 
(R_o )v(R_o ).$ Here $n_I (R_o )$ is the value at $R_o $ of another primary 
dependent variable, the ion density at $r,$ $n_I (r).$ Steady-state 
conditions are assumed to prevail in the model, and since it is also 
postulated that \textit{all} of the ions escape into interstellar space, an effective 
equation of continuity must exist. One therefore would have 
\begin{equation}
\label{eq9}
4\pi r^2n_I (r)v(r)=K_2 .
\end{equation}
The third primary dependent variable is $\phi _2 (r),$ the flux at $r$ of the 
monochromatic, radially inwardly propagating laser wave at $\nu _2 \equiv 
\nu _o .$ The growth of $\phi _2 (r)$ occurs due to stimulated Rayleigh 
scattering, and can be represented by the equation
\begin{equation}
\label{eq10}
-\frac{d\phi _2 (r)}{dr}=\sigma (r)n_I (r)\phi _1 (r)\phi _2 (r).
\end{equation}
With this equation, two additional dependent variables, $\phi _1 (r)$ and 
$\sigma (r),$ appear to have been introduced. However, it is straightforward 
to write down the functional dependence on $r$ of one of these variables, and 
the other variable can easily be expressed in terms of the previously 
introduced dependent variable $v(r).$ Both these variables will now be 
discussed. 

The function $\phi _1 (r)$ is the star's continuum flux at position $r$ that 
is effective in pumping the stimulated Rayleigh scattering process occurring 
at that point. Basically, it represents the continuum light at $r$ contained 
within a certain fixed narrow band width $\Delta \nu _n .$ This band width 
$\Delta \nu _n $ should approximately correspond to the spectral line width 
of the inwardly propagating $\nu _2 $ laser radiation. Initially, it can be 
regarded as an adjustable parameter. On the basis of the proposed model, one 
can write $\phi _1 (r)=K_1 /r^2,$ with $K_1 $ being a constant easily 
determinable from the emissive properties of the star generating the solar 
wind. The reason why it is possible to write $\phi _1 (r)$ as a simple 
inverse square function of $r$ follows from the discussion given in Sec. 
III. At any radius $r$, the only pump light that is effectively absorbed 
occurs in a narrow band width $\Delta \nu _n $ that is spectrally located at 
$\simeq \nu _o +2\nu _o (v/c).$ Thus, nowhere in the solar wind is the 
continuum pump light depleted by ions located at positions with smaller $r$ 
values. 

The function $\sigma (r)$ is an effective two-photon cross-section for the 
stimulated Rayleigh process. Its functional dependence on $r$ is given by 
\begin{eqnarray}
\label{eq11}
\sigma (r)&\simeq& \frac{\sigma _o }{\left\{ {\nu _1 (r)\left[ 
{1-\frac{v(r)}{c}} \right]-\nu _o } \right\}^2}\nonumber \\
&\simeq& \frac{\sigma _o 
}{\left\{ {\left[ {\nu _o +2\nu _o \frac{v(r)}{c}} \right]\left[ 
{1-\frac{v(r)}{c}} \right]-\nu _o } \right\}^2}\\
&\simeq& \frac{\sigma _o 
c^2}{\nu _o^2 v^2(r)}\nonumber,
\end{eqnarray}
this equation simply reflecting the fact that excitation is occurring on the 
wing of a lifetime broadened Lorentzian resonance profile. The constant 
$\sigma _o $ can be determined from the radiative properties of the P Cygni 
resonance transition. 

A third independent equation relates two of the three inherently dependent 
variables to each other. The radial velocity $v(r)$ of each ion located in 
the volume between $r$ and $r+dr$ will be increased by an amount $\sigma 
(r)\phi _1 (r)\phi _2 (r)(\Delta v)$ per second. Since it takes a time 
$dr/v(r)$ for an ion to reach $r+dr,$ one can write 
\begin{equation}
\label{eq12}
v(r)\frac{dv(r)}{dr}=\sigma (r)\phi _1 (r)\phi _2 (r)(\Delta v).
\end{equation}
From Eqs. (\ref{eq9}), (\ref{eq10}), and (\ref{eq12}), it appears that obtaining a solution to the 
system of equations here discussed amounts to solving the following two 
coupled equations: 
\begin{equation}
\label{eq13}
-\frac{d\phi _2 (r)}{dr}=\left[ {\frac{\sigma _o c^2K_2 K_1 }{4\pi \nu _o^2 
}} \right]\frac{\phi _2 (r)}{r^4v^3(r)}
\end{equation}
and 
\begin{equation}
\label{eq14}
v^3(r)\frac{dv(r)}{dr}=\left[ {\frac{(\Delta v)\sigma _o c^2K_1 }{\nu _o^2 
}} \right]\frac{\phi _2 (r)}{r^2}.
\end{equation}
All the constants appearing in brackets in these two equations are in 
principle known for transitions having P Cygni line shapes in stars such as 
$\zeta $ Pup. It would be interesting to see if numerical solutions can 
indeed reveal a threshold type of behavior, with parameter values in certain 
ranges resulting in the occurrence of ``lasing'' in the star's solar wind. 

\section{Possible spectroscopic consequences of the proposed model}

\subsection{The fluorescence component of a P Cygni profile}

As has been indicated above, a signature feature of the proposed nonlinear P 
Cygni model is an intense monochromatic $\nu _2 $ laser wave that radially 
impinges upon the star's photosphere. Since this wave would be directly 
unobservable in any line of sight, one should therefore consider what kinds 
of observable \textit{secondary} effects might result from its presence. 

One predictable effect would be \textit{fluorescence} seen in the line-of-sight that is emitted 
by P Cygni ions moving in the solar wind. As discussed in various texts 
($e. g.$ Loudon \cite{loudon}), when there is no collision broadening, and the excitation 
intensity is weak, the Lorentzian cross-section for linear resonant light 
scattering by a two-level atom exclusively applies to the elastic ($i. e.$ 
Rayleigh) component. The probability of inelastic scattering is zero, $i. e.$ no 
real linear excitation of the atom occurs. The situation becomes entirely 
different, however, when intense excitation (as produced by a laser beam, 
for example) is present. Then real excitation of the upper level of the atom 
can result, even when the exciting light is non resonant ($c. f.$ Knight {\&} 
Milonni \cite{knight}). One thus could expect ions in the solar wind to undergo real 
transitions to the upper level of the P Cygni resonance, driven by an 
intense inwardly propagating laser wave at $\nu _2 .$ Such ions would then 
fluoresce. At first glance, it might appear that the fluorescence occurring 
at $\nu >\nu _o $ would be absorbed in pumping the stimulated Rayleigh 
scattering processes that occur in the solar wind. However, it can be 
readily shown that this cannot be the case. Consider an ion in the solar 
wind moving with velocity $v$ in a direction towards Earth. If this ion 
becomes inelastically excited by the $\nu _2 $ laser wave, the fluorescence 
it immediately would radiate towards Earth would occur at the frequency $\nu 
_o +\nu _o (v/c).$ However, it was shown in Sec. III that the continuum pump 
light necessary to make the same ion undergo stimulated Rayleigh scattering 
occurs at frequency $\nu _o +2\nu _o (v/c).$ \textit{Thus, the fluorescence emitted by the ion towards Earth} \textit{cannot} \textit{be} \textit{nonlinearly absorbed anywhere in the solar wind}. (In principle, it could be 
linearly absorbed to some extent by surrounding ions in the solar wind, but 
in the proposed nonlinear model this process is being entirely neglected.) 
Fluorescence in the line-of-sight occurring at $\nu <\nu _o $ will neither 
be linearly nor nonlinearly absorbed.

In view of the deduction made in the last paragraph, it would appear that, 
on the basis of the proposed nonlinear model, a rough replica of a typical P 
Cygni profile could be conceptually constructed in the following manner. 
Start with the basic continuum level of the star. Subtract from this the 
blueshifted light that is absorbed in pumping stimulated Rayleigh 
scattering. To this depleted continuum then simply add the fluorescence 
intensity radiated towards Earth by all the ions in the solar wind. Such 
fluorescence will generally be much stronger for $\nu >\nu _o $ than for $\nu 
<\nu _o ,$ due to occultation by the star. Via this conceptual construction, 
a simple explanation is therefore provided for a characteristic observed 
feature of P Cygni profiles, namely, that the strongest attenuation of the 
continuum appears to occur at frequencies that are significantly offset from 
$\nu _o .$ In reality, strong attenuation of the continuum may also be 
occurring much closer to $\nu _o ,$ but this is masked by the presence of a 
strong fluorescence signal. 

From the fluorescence mechanism here proposed, it also follows that the 
reddest fluorescent emission observed should have an absolute value 
frequency offset from $\nu _o $ no greater than half that of the bluest P 
Cygni absorption. In most P Cygni spectra this seems approximately to be the 
case. However, it is perhaps well to note here that fluorescence appears not 
to be as intrinsic a feature of P Cygni profiles as is extended blueshifted 
absorption. For the P Cygni spectra observed in Morton (\cite{morton76}), it was noted 
that the equivalent width of the emission component is always significantly 
less than that of the absorption component. In the P Cygni profile of N III 
in $\zeta $ Pup, in fact, the fluorescence component is completely absent. 
It was suggested in Morton (\cite{morton76}) that this might be the result of 
absorption occurring close enough to the stellar surface to hide the 
emission via occultation by the star. To obtain even a rough estimate of the 
expected fluorescence profile based upon the nonlinear model would require 
at least semi-quantitative solutions of equations such as those discussed in 
Sect. 4 to be made. From the present section, it would appear that one 
should modify those equations by including a loss term for the variable 
$\phi _2 (r),$ in order to represent strong fluorescence being excited. 

\subsection{Two-photon absorption bands}

In principle, one could seek to directly substantiate the proposed nonlinear 
model by clearly identifying in a P Cygni spectrum Doppler broadened 
absorption bands representing resonantly enhanced two-photon absorption of 
continuum light from the star by various ion species present in its 
photosphere, or just outside. Such two-photon absorptions could in principle 
be induced if powerful, monochromatic beams at P Cygni ion rest frame 
frequencies $\nu _o $ were indeed incident upon the star's photosphere, as 
the nonlinear model presupposes. For such two-photon absorptions to be 
strong enough to be detectable, near resonances would have to exist between 
some of the P Cygni rest frame frequencies and some of the transitions of 
the two-photon-absorbing ions. 

\begin{figure}
\resizebox{\hsize}{!}{\includegraphics{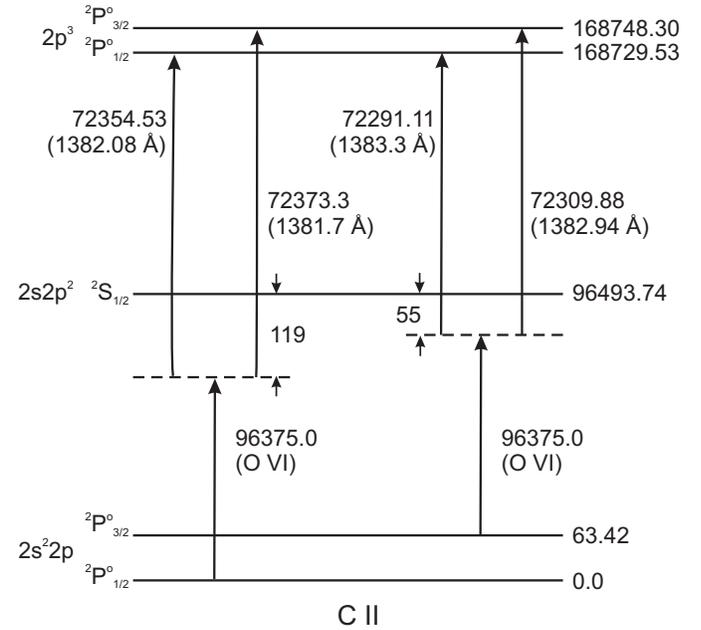}}
\caption{C II partial energy level diagram showing two-photon 
absorption transitions that are likely to be induced by application of 
strong monochromatic coherent radiation at 96375 cm$^{-1}$, the frequency of 
the O VI resonance at 1038 {\AA}.}
\label{fig3}
\end{figure}

The authors have attempted to identify such two-photon absorption bands in 
the case of the P Cygni star $\zeta $ Pup. With use of the NIST Atomic 
Spectra Database web site, spectral intervals extending $\sim $ 200 cm$^{-1}$ 
on either side of the rest frame frequencies $\nu _o $ of the strongest P 
Cygni profiles shown in Morton (\cite{morton76}) were examined for likely absorbing-ion 
candidates, but no convincing ones were found. The spectral intervals 
examined were as follows: C III (977 $\pm $ 2{\AA}); C IV (1551 $\pm $ 5{\AA}, 
1548 $\pm $ 5{\AA}); N V (1243 $\pm $ 3.2{\AA}, 1239 $\pm $ 3.2{\AA}); Si IV 
(1403 $\pm $ 4{\AA}, 1394 $\pm $ 4{\AA}); N III (992 $\pm $ 3{\AA}, 990 $\pm 
$ 3{\AA}); S VI (945 $\pm $ 5{\AA}, 933 $\pm $ 5{\AA}); and O VI (1032 $\pm 
$ 5{\AA}, 1038 $\pm $ 5{\AA}). The most promising coincidence found was between 
the wavelength (1038 {\AA}) of one of the O VI doublet components and a 
strongly allowed C II ion transition originating from a level only 63 
cm$^{-1}$ above its ground state (Fig.~\ref{fig3}). The difference between the O VI 
resonance and C II transition frequency is in this case only 55 cm$^{-1}$. 
With this scheme (and with another somewhat less resonant one also shown in 
Fig.~\ref{fig3}), one predicts two-photon absorption bands to exist at 1381.7 {\AA}, 
1382.08 {\AA}, 1382.94 {\AA}, and 1383.3 {\AA}, with the latter two being 
stronger than the former two. 

On the basis of several high resolution VUV \textit{Copernicus } scans of $\zeta $ Pup recorded 
in 1973-1975, interstellar absorption lines seen in the line-of-sight to 
this star were identified and tabulated in Morton (\cite{morton78}). This same 
publication also contains a table of 52 lines towards $\zeta $ Pup that were 
unidentifiable in 1978. In 1991, Morton published a large compendium (Morton 
\cite{morton91}) of interstellar absorption lines based upon data obtained from many 
space objects with use of the \textit{Hubble Space Telescope}. Included in this finding list was an updated 
version of the original unidentified lines list of Morton (\cite{morton78}), the number 
of such lines towards $\zeta $ Pup by 1991 having been reduced to 42, 
largely as a result of the discovery of seven new Fe II lines that were not 
known in 1978. Still remaining unassigned in 1991 were four lines in $\zeta 
$ Pup occurring as an isolated group at 1382.305 {\AA} (8.9), 1382.593 {\AA} 
(22.1), 1383.205 {\AA} (30.8), and 1383.720 {\AA} (18.6) (equivalent widths 
in m{\AA} in parentheses). One sees that the four unidentified lines in 
$\zeta $ Pup are tantalizingly close to the four predicted two-photon 
absorption bands. However, despite the existence of these intriguingly close 
wavelength matches, it would be apparently difficult to argue that C II 
would not be fully ionized in the photosphere of a star as hot as $\zeta $ 
Pup. Strong C II absorptions are seen in the $\zeta $ Pup spectrum, but 
astronomers believe that these represent interstellar lines. 

It is of interest here to comprehend roughly what the maximum intensity of 
the monochromatic $\nu _2 $ wave could be as it impinges upon a star's 
photosphere. We here consider the case where $\nu _2 $ corresponds to the O 
VI P Cygni resonance at 1032 {\AA} in $\zeta $ Pup. In this star, the 
corresponding blueshifted absorption region extends to about 1020 {\AA}. 
Under the assumption that roughly 80{\%} of the continuum photons emitted by 
the star between 1020 {\AA} and 1032 {\AA} are nonlinearly absorbed and 
entirely converted to photons of the inwardly propagating monochromatic wave 
at $\nu _2 $, it follows that the intensity of the latter would be roughly 
200 kW/cm$^{2}$ as it impinges upon the star's photosphere. In this 
estimate it is assumed that the temperature of $\zeta $ Pup is 50,000K, and 
that this star emits as a perfect blackbody. It is also here assumed that no 
loss of $\nu _2 $ wave intensity occurs as a result of fluorescence 
excitation. As outlined in Sect. 5a, one actually expects such loss to be 
quite large. 

\section{Linear vs. nonlinear P Cygni mechanisms in luminous stars}

As noted above, current theoretical models for line driven winds of luminous 
hot stars are all based upon the effects of \textit{linear} resonant absorption of 
continuum light emitted by such stars. In such models, the essential 
physical processes that occur are in many ways similar to those that 
characterize the nonlinear model described in Sect. 3. (Chapter 8 of Lamers 
{\&} Cassinelli (\cite{lamers}) is a good exposition of the currently accepted linear 
theory of line driven winds.) For example, it is easy to demonstrate that 
the outwardly directed momentum transfer to an atom or ion in the solar wind 
in a linear scattering event is $h\nu _o /c-i.e.$ \textit{half} the value occurring in the 
unit step of the nonlinear process. In the linear theory, just as in the 
nonlinear case, the Doppler shifts of atoms or ions moving in the outer 
portions of a solar wind in principle allow the atoms to absorb undiminished 
continuum photons in their line transitions. However, the light that 
effectively drives the acceleration of an atom moving with a given radial 
velocity $v$ in the case of the nonlinear mechanism is blueshifted from $\nu 
_o $ \textit{twice }as much as in the linear case. This implies that occurrence of the 
former mechanism ($i. e.$ stimulated Rayleigh scattering) would have already 
strongly depleted the supply of continuum photons necessary to drive the 
linear process everywhere in the solar wind. Since the stimulated Rayleigh 
scattering process should occur at its highest rate very close to the 
photosphere of the illuminating star, it is even possible that in very 
bright stars it completely prevents linear scattering from being an 
effective acceleration mechanism. 

\begin{acknowledgements}P.P.S. is grateful to IBM Research management for having 
allowed him, as an IBM Fellow Emeritus, the use of office and library 
facilities for the past three-and-a-half years.
\end{acknowledgements}

\end{document}